\documentclass{l4dc2024}
\usepackage{times}
\newtheorem{Assumption}{Assumption}
\newtheorem{Remark}{Remark}

\usepackage{amsmath}               
\usepackage{algpseudocode}
\usepackage{algorithm}
\usepackage{algpseudocode}
\usepackage{xcolor}
\usepackage{algcompatible}

\SetKwInput{KwInput}{Input}                
\SetKwInput{KwOutput}{Output}              

\title[Data-Driven Safety Filter]{From Raw Data to Safety: Reducing Conservatism by Set Expansion}

\author{%
 \Name{Mohammad {Bajelani}} \Email{mohammad.bajelani@ubc.ca}\\
 \Name{Klaske {van Heusden}} \Email{klaske.vanheusden@ubc.ca }\\
 \addr School of Engineering\\
       University of British Columbia,\\
       3333 University Way, Kelowna, BC Canada V1V 1V7
}

\begin{document}
\maketitle

\editor{---}

\begin{abstract}
In response to safety concerns associated with learning-based algorithms, safety filters have been proposed as a modular technique. Generally, these filters heavily rely on the system's model, which is contradictory if they are intended to enhance a data-driven or end-to-end learning solution. This paper extends our previous work, a purely Data-Driven Safety Filter (DDSF) based on Willems' lemma, to an extremely short-sighted and non-conservative solution. Specifically, we propose online and offline sample-based methods to expand the safe set of DDSF and reduce its conservatism. Since this method is defined in an input-output framework, it can systematically handle both unknown and time-delay LTI systems using only one single batch of data. To evaluate its performance, we apply the proposed method to a time-delay system under various settings. The simulation results validate the effectiveness of the set expansion algorithm in generating a notably large input-output safe set, resulting in safety filters that are not conservative, even with an extremely short prediction horizon.
\end{abstract}

\section{Introduction}
Nowadays, learning-based and data-driven approaches outperform traditional controllers in terms of superior performance without the assistance of expert knowledge, especially in addressing hard-to-model problems \citep{brunke2022safe}. Despite these advancements, the integration of such methods poses challenges in ensuring real-world safety during the learning process \citep{hewing2020learning}. Generally, the safety criteria are defined by input-output constraints, such as restricted torque and a pre-specified workspace in a robot manipulator \citep{wabersich2021predictive} or the load factor in flight envelope for an aircraft \citep{hsu2023towards}. To respect these constraints during the learning process, safety filters have been proposed within the control community as a modular technique to ensure safety irrespective of the learning method, whether provided by a human or a reinforcement learning agent \citep{wabersich2018linear}. The main idea behind these safety filters is the mapping of potentially unsafe learning inputs to the nearest safe learning inputs while satisfying specific criteria, such as ensuring the system's safety, i.e., forcing the system's trajectory to live in some invariant sets, over a finite or infinite time horizon.

Currently, there are three main streams to designing safety filters \citep{wabersich2023data} which all rely on accurate and explicit models. The utilization of Control Barrier Functions (CBFs) allows for the creation of safety filters capable of gently adjusting a specified input control signal as the system approaches the boundary of a control invariant set \citep{ames2019control}. Despite CBFs' proven effectiveness in applications requiring swift responses \citep{8460471}, a notable challenge arises in formulating barrier functions or ensuring the feasibility of QP-based CBFs in the presence of input constraints \citep{molnar2023safety}. Hamilton-Jacobi (HJ) reachability analysis is a tool for ensuring system safety by defining reachable sets in the state space. These sets represent areas where specific goals can be achieved or safety conditions are satisfied \citep{bansal2017hamilton}. However, a major drawback is its computational complexity, especially for higher-dimensional systems \citep{herbert2021scalable}. Lastly, Model Predictive Safety Filters (MPSFs) use an explicit predictive model to evaluate the safety of incoming learning inputs and then map them to the closest safe actions by constructing the backup trajectories \citep{tearle2021predictive}. The primary challenges associated with this approach are the substantial computational requirements for online processing and the inherent complexity of robust design. For a comprehensive examination of their advantages and disadvantages, as well as data-driven approaches for the modeling component, see the recent survey by \citep{wabersich2023data}. 

It is critical to underscore that these methods all depend on explicit models represented in state space, constituting a drawback for describing the uncertainty of the (multi-step) prediction \citep{kohler2022state}, as determining non-conservative bounds for state-space predictions is challenging. Furthermore, if these safety methods are intended to guarantee the safety of end-to-end learning algorithms, they should bypass the system identification process; otherwise, it is counterproductive to adopt an online learning algorithm to learn the system for which an accurate explicit model is available. In the indirect approach (system identification \& control), uncertainty from the data must be transmitted through the system identification process, potentially causing mismatches with preferred uncertainty quantification for control design. In contrast, direct approaches (directly from data to control) seamlessly integrate uncertainty into control design, eliminating the need for complex uncertainty propagation and offering a transparent and effective strategy in data-driven control \citep{10317633}.

This paper proposes an entirely data-driven approach to determine safe terminal sets in the input-output framework. Combined with the recently introduced data-driven safety filter (DDSF) \citep{bajelani2023data}, this work achieves a non-conservative short-sighted safety filter law directly from raw data. It should be noted that, if an explicit model is available, terminal safe sets can be calculated. However, results for computation of terminal sets in the data-driven framework are limited to (trivial) solutions like an equilibrium point \citep{berberich2020data,BERBERICH20206923} or knowledge of the system's lag \citep{berberich2021design}. Motivated by \citep{rosolia2017learning_1,rosolia2017learning_2}, we extend set expansion to the input-output framework, resulting in a solution that requires only a single batch of data. For readability, the methods in this paper are derived assuming noise-free measurements are available. While the extension to data-driven predictive control that takes noise into account can be applied directly to the DDSF, the impact of noise or unmodeled dynamics on terminal sets has received little attention. The proposed data-driven set expansion offers a direct connection between noise and computed sets, extending the data-driven framework's transparency to terminal sets.

The remaining sections of the paper are structured as follows: Section 2 covers preliminary material for the data-driven predictive safety filter based on Willems' Lemma. Moving to Section 3, the formulation of DDSF with a sampled safe set is presented, demonstrating how the final set can be expanded using the input-output trajectories of the system or backup trajectories calculated by the DDSF. Section 4 provides simulation results for various settings, focusing on a second-order system with an unknown time delay by online and offline set expansion algorithms. Lastly, Section 5 comprises of the discussion, concluding remarks, and potential avenues for future exploration.

\newpage
\section{Background}
Consider a discrete-time LTI system as follows:
\begin{equation} \label{equ: true system}
    x(t+1) =A x(t)+B u(t-\tau_{d}), \quad
    y(t)=C x(t)+D u(t-\tau_{d}),
\end{equation}
where $A \in \mathbb{R}^{n \times n}, B \in \mathbb{R}^{n \times m}, C \in \mathbb{R}^{p \times n}, D \in \mathbb{R}^{p \times m}$ are unknown matrices. The input, state, and output vectors are respectively denoted by $u(t) \in \mathbb{R}^m, x(t) \in \mathbb{R}^n, y(t) \in \mathbb{R}^p$ at time $t \in \mathbb{Z}_{\geq 0}$. Note that in addition to the matrices, the order of the system $n$ and the time delay $\tau_{d} \in \mathbb{Z}_{\geq 0}$ are unknown. It is also assumed that the pair of $(A,B)$ and $(A,C)$ are controllable and observable. The time-delay system (\ref{equ: true system}) can be represented as an augmented free-delay system (\ref{equ: augmented time-delay-free system}) with additional internal states as follows,
\begin{subequations} \label{equ: augmented time-delay-free system}
    \begin{align}
        & \begin{matrix}
            x_a(t+1) = A_a x_a(t)+B_a u(t),\quad y(t) = C_a x_a(t) +D u(t),
        \end{matrix}\\
        & x_a (t) := \begin{bmatrix}  
        x(t) \\
        d(t)
        \end{bmatrix}, \ d(t) := \begin{bmatrix}  
        u(t-\tau_{d}) \\ u(t-\tau_{d}+1) \\ \vdots \\ u(t-1)
        \end{bmatrix},\\
        &A_a := \begin{bmatrix}
        A_{n \times n} \ B_{m \times n} \ 0_{(n \times (k_d-1)) \times m} \\
        0_{(m \times (k_d-1)) \times n+m} \  \ I_{(m \times (k_d-1)) \times (m \times (k_d-1))} \\
        0_{1 \times (n+m+k_d)}
        \end{bmatrix} ,\\
        & B_a := \begin{bmatrix}
        0_{(m \times (k_d-1)) \times m} \\   I_{ m \times  m}    
        \end{bmatrix},
        C_a := \begin{bmatrix}
        C_{p \times n}   \\ 0_{(1 \times  m) \times k_d}    
    \end{bmatrix}.
    \end{align}     
\end{subequations}
In addition, input-output safety is represented by a set of polytopes in the form of 
\begin{subequations} \label{equ: I-O constranits} 
    \begin{align}   
        u(t) \in \mathcal{U} &:= \{ u \in \mathbb{R}^m \mid A_u u < b_u, A_u \in \mathbb{R}^{n_u \times m}, b_u \in \mathbb{R}^{n_u} \}, \\
        \quad y(t) \in \mathcal{Y} &:= \{ y \in \mathbb{R}^p \mid A_y y < b_y, A_y \in \mathbb{R}^{n_p \times p}, b_y \in \mathbb{R}^{n_p}\},
    \end{align}   
\end{subequations} 
for all $t \in \mathbb{Z}_{\geq 0}$. The primary objective of a safety filter is to modify any (potentially) unsafe learning inputs, $u_{l} \in \mathbb{R}^m$, as minimally as possible while ensuring that the system (\ref{equ: true system}) remains within the defined bounds $(\mathcal{U},\mathcal{Y})$ defined in (\ref{equ: I-O constranits}) for infinite time. Throughout this paper, it is assumed that a single noise-free input-output measured trajectory of the system (\ref{equ: true system}) in the form of (\ref{equ: Input-output_data}) is available, which is sufficiently exciting in the sense of Definition (\ref{Def: PE condition}).
\begin{equation} \label{equ: Input-output_data}
        u^d_{[0,{N_0-1}]} = [{u_0}^\top, \hdots, {u_{N_0-1}}^\top]^\top, \quad
        y^d_{[0,{N_0-1}]} = [{y_0}^\top, \hdots, {y_{N_0-1}}^\top]^\top.
\end{equation}
To build an implicit model from the pre-recorded trajectory (\ref{equ: Input-output_data}), we reshape these two sequences into the form of a Hankel matrix shown by (\ref{equ: Hankel}). These matrices are used to provide an implicit model to represent the input-output behavior of the system (\ref{equ: true system}).   
\begin{equation} \label{equ: Hankel}
        H_L(u) = \begin{bmatrix}
            u_0 & u_1 & \ldots & u_{N_0-L} \\
            u_1 & u_2 & \cdots & u_{N_0-L+1} \\
            \vdots & \vdots & \ddots & \vdots \\
            u_{L-1} & u_L & \ldots & u_{N_0-1}
        \end{bmatrix},  
        H_L(y) = \begin{bmatrix}
            y_0 & y_1 & \ldots & y_{N_0-L} \\
            y_1 & y_2 & \cdots & y_{N_0-L+1} \\
            \vdots & \vdots & \ddots & \vdots \\
            y_{L-1} & y_L & \ldots & y_{N_0-1} 
        \end{bmatrix},
\end{equation}
where $H_L(u) \in \mathbb{R}^{(mL)\times(N_0-L+1)}$ and $H_L(y) \in \mathbb{R}^{(pL)\times(N_0-L+1)}$. The raw data stored in the stacked Hankel matrices $[H_L(u) H_L(y)]$ is used to directly parameterize any input-output trajectory of the system (\ref{equ: true system}) using the fundamental Lemma proposed by Jan Willems in \citep{willems2005note}.
\begin{definition}[System's Lag] \label{definition: lag} 
$\underline{l}=l(A, C)$ denotes the lag of the system (\ref{equ: true system}), which is the smallest integer that can make the observability matrix full rank
        \begin{equation}
        O_l(A, C):=\left(C, C A, \ldots, C A^{l-1}\right).
        \end{equation}
\end{definition}
\begin{definition}[Extended State \citep{berberich2021design}] \label{definition: Extended State}
For some integers $T_{\text{ini}}>\underline{l}~$, the extended state $\xi_t$ at time $t$ is defined as follows
    \begin{align}\label{eq:extended_state}
    \xi_t := \begin{bmatrix}u_{[t-T_{\text{ini}},t-1]}\\y_{[t-T_{\text{ini}},t-1]}\end{bmatrix} \in \mathbb{R}^{(m+p)T_{\text{ini}} \times 1} .
    \end{align}
where $u_{[t-T_{\text{ini}},t-1]}$ and $y_{[t-T_{\text{ini}},t-1]}$ denote the last $T_{\text{ini}}$ input and output measurements, respectively.
\end{definition}
Utilizing sufficiently long past input-output measurements, known as the extended state, enables the determination of the internal state of the underlying and unknown system \citep{coulson2019data}. The minimally required length of the extended state is dependent on the system's lag (Definition (\ref{definition: lag})). The minimum length of $L$, representing the number of rows in Hankel matrices, is determined by the order of persistent excitation of input $u$ as defined in Definition (\ref{Def: PE condition}).
\begin{definition}[Persistently Excitation \citep{berberich2020data}] \label{Def: PE condition}
     Let the Hankel matrix's rank be $rank(H_L(u)) = mL$, then $u \in \mathbb{R}^m$ represents a persistently exciting signal of order $L$.
\end{definition}
Behavioral system theory differs from classical system theory in its approach to viewing dynamical systems. It focuses on systems as input-output trajectories defined in the signal subspaces \citep{markovsky2006exact}. It has been demonstrated that if a finite-time trajectory of an LTI system is available and the input is persistently exciting, all the trajectories can be parameterized linearly by combining the columns of the Hankel matrix. This concept is known as the fundamental lemma introduced by Jan Willems \citep{willems2005note}. It also enables us to predict input-output trajectories of the true system (\ref{equ: true system}) directly from raw data without requiring an explicit model. By using the implicit model derived from the fundamental lemma, we avoid the inevitable under- or over-modeling associated with explicit methods. For a recent overview of implicit vs explicit perspectives, see \citep{10317633,markovsky2021behavioral}.
\begin{theorem}[Fundamental Lemma \citep{berberich2020robust}] \label{Fundamental Lemma}
    Let $u^d$ be persistently exciting of order $L+n$, and ${\{u^d_k},{y^d_k\}}_{k=0}^{k=N_{0}-1}$ a trajectory of $G$. Then, ${\{\bar{u}_k},{{\bar{y}_k}\}}_{k=0}^{k=N_{0}-1}$ is a trajectory of $G$ if and only if there exists $\alpha \in \mathbb{R}^{N_{0}-L+1}$ such that
\begin{equation} \label{BST model}
    \left[\begin{array}{l}
    H_L\left(u^d\right) \\
    H_L\left(y^d\right)
    \end{array}\right] \alpha=\left[\begin{array}{l}
    \bar{u} \\
    \bar{y}
    \end{array}\right] .  
\end{equation}
\end{theorem}
\begin{Remark}
The application of model-based safety filters poses challenges for systems with unknown time delays. The input-output framework in behavioral system theory inherently addresses this, emphasizing essential input-output measurements. Given the equivalent augmented system representation that treats the delayed inputs as internal states, denoted by $d(t)$ in (\ref{equ: augmented time-delay-free system}), the input-output behavior of the implicit model provided in Theorem (\ref{Fundamental Lemma}) and systems (\ref{equ: true system}-\ref{equ: augmented time-delay-free system}) are equivalent.
\end{Remark}
\begin{definition} [Invariant Set] \label{def: Invariant Set.}
    A set $\Xi \subseteq \mathbb{R}^{(mT_{\text{ini}}+pT_{\text{ini}})}$ is said to be a control invariant set for the system (\ref{equ: true system}) subjected to constraints in (\ref{equ: I-O constranits}) from the input-output perspective, if 
    $$\xi(t) \in \Xi \quad \Rightarrow \quad \exists u(t) \in \mathcal{U} \text { such that } \xi(t+1) \in \Xi, \quad \forall t \in \mathbb{Z}_{\geq t}$$
\end{definition}
\begin{definition} [Equilibrium Point] \label{def: Equilibrium point.}
    $\left(u^s, y^s\right) \in$ $\mathbb{R}^{m+p}$ is an equilibrium point of system (\ref{equ: true system}), if the extended state $\xi^s$ defined by sequence $\left\{{u}_k, {y}_k\right\}_{k=0}^{k=T_{\text{ini}}-1}$ with $\left({u}_k, {y}_k\right)=\left(u^s, y^s\right)$.
\end{definition} 
Note that the equilibrium point defined in Definition (\ref{def: Equilibrium point.}) represents a special case of an invariant set as defined in Definition (\ref{def: Invariant Set.}). In essence, if the system's input and output can be maintained at the equilibrium point for more than $T_{\text{ini}}$ steps, all underlying states of the system (\ref{equ: true system}) are fixed at their equilibrium values.

\section{Data-Driven Safety Filter with Sampled Terminal Sets}

In this section, an extended version of DDSF based on behavioral system theory, as proposed in \citep{bajelani2023data}, is introduced. More specifically, the prior method assumed that the terminal safe set should align with the system's equilibrium point which results in an excessively conservative system, particularly when dealing with a short prediction horizon. As a consequence, the safety filter prevents the learning process by over-correcting the learning input. To tackle this challenge, this paper proposes two sample-based approaches motivated by \citep{rosolia2017learning_2, wabersich2018linear}. The first approach is designed for online implementation which only needs new experiment data, while the second one allows the offline calculation of data-driven safe sets using a single dataset. The sampled safe sets are represented in (\ref{eq:SS_online}) for the online approach at time $t$ and (\ref{eq:SS_offline}) for the offline approach at iteration $K$, respectively.
\begin{subequations} 
    \begin{gather}   \label{eq:SS_online}
        {\bar\Xi}_f^t = \textrm{}\text{Conv}(\bigcup_{k=0}^{t-1} \xi_k)\\\label{eq:SS_offline}
        {\bar\Xi}_f^K = \textrm{}\text{Conv}(\bigcup_{l=0}^{K-1}\bigcup_{j=0}^{N-1} \bar{\xi}^l_j).
    \end{gather}   
\end{subequations} 
where $\text{Conv(.)}$ is the convex hull, $\xi_k$ is the extended state at time $t=k$, $\bar{\xi}^l_j$ is the extended state provided by the $l^\text{th}$ element of backup trajectory at iteration $j$, $K$ is the number of iteration, and $N$ is the prediction horizon length. Using the definition of the convex hull, the terminal sets (\ref{eq:SS_online}-\ref{eq:SS_offline}) can be written in the form of (\ref{eq:CS_online}-\ref{eq:CS_offline}) for the online and offline approaches,
\begin{subequations} 
    \begin{gather}   \label{eq:CS_online}
        {\bar\Xi}_f^t = \Big\{ \sum_{t=0}^{t-1} \alpha_{t} \xi_t : \alpha_t \geq 0, \sum_{t=0}^{t-1} \alpha_{t} = 1\}, \\\label{eq:CS_offline}
        {\bar\Xi}_f^K = \Big\{ \sum_{l=0}^{K-1}\sum_{j=0}^{N-1} \alpha^l_j \bar{\xi}^l_j : \alpha^l_j \geq 0, \sum_{j=0}^{N-1} \alpha^l_j = 1\}.
    \end{gather}   
\end{subequations} 
\begin{Remark}
    The main idea behind the sampled safe set is that each system trajectory is an invariant set, and the union of safe trajectory is an invariant set. For linear systems, since the feasible is a convex set and the input-output constraints $(\mathcal{U}, \mathcal{Y})$ are polytopes, then the safe set is a convex set. Thus, the safe sets can be recovered by sampling extended state or backup trajectory convex hulls.
\end{Remark}
If more data or backup trajectories are collected, the final set $\Xi_f$ can be extended (\ref{eq:SS_online}-\ref{eq:SS_offline}). Hence, the size of the sampled safe set does not decrease with the addition of more data, indicating that if the first safe set is not empty, then the subsequent ones are also not empty as described as follows,
\begin{subequations} \label{subsets}
    \begin{align} 
        {\bar\Xi}_f^{t-1} \subseteq {\bar\Xi}_f^t, \\
        {\bar\Xi}_f^{K-1} \subseteq {\bar\Xi}_f^K.
    \end{align}   
\end{subequations}
\begin{figure}[t]
    \begin{minipage}{0.48\textwidth}
        \begin{algorithm}[H] 
        \caption{Set Expansion (Online)}\label{algorithm: DDSF_online}
        \KwInput{$H_L(u^d)$, $H_L(y^d)$, $N_p$, $T_{\text{ini}}$, and $\Xi^{t=0}$}
        \KwOutput{$\Xi_f$}
        initialization\;
        \While{true}{
        Solve problem (\ref{eq:DDSF}) for $u_l(t)$\\
        Apply the safe input to the system (\ref{equ: true system}) and update the initial condition\\
        Expand the safe set $\Xi^{t}$ using (\ref{eq:SS_online})\\
        \If{$\Xi^{t-1}\approx\Xi^{t}$}{break the while loop}
        $t\Rightarrow t+1$}
        \end{algorithm}
    \end{minipage}%
    \hfill
    \begin{minipage}{0.48\textwidth} 
        \begin{algorithm}[H] 
        \caption{Set Expansion (Offline)} \label{algorithm: DDSF_offline}
        \KwInput{$H_L(u^d)$, $H_L(y^d)$, $N_p$, $T_{\text{ini}}$, and $\Xi^{l=0}$}
        \KwOutput{$\Xi_f$}
        initialization\;
        \While{true}{
        Solve problem (\ref{eq:DDSF}) for $u_l(l)$\\
        Choose an element of backup trajectory as the initial condition\\
        Expand the safe set $\Xi^l$ using (\ref{eq:SS_offline})\\
        \If{$\Xi^{l-1}\approx\Xi^{l}$}{break the while loop}
        $l\Rightarrow l+1$}
        \end{algorithm}
    \end{minipage}
\end{figure}
A visualization of the evolution of these sets for the online approach is shown in Figure (\ref{fig: Extended_Sample_Safe_Set}). By comparing Figures (\ref{fig: Extended_Sample_Safe_Set}A) and (\ref{fig: Extended_Sample_Safe_Set}B), it can be seen that when the final safe set is expanded, conservatism is reduced and performance is improved. This allows the learning algorithm to explore the space more and infer less with its learning process. A significant advantage of the online approach (\ref{algorithm: DDSF_online}) is the possibility of collecting a sample safe set after each time step without collecting a complete iteration as proposed in \citep{rosolia2017learning_1}. Furthermore, the offline approach enables us to expand the safe set only using one batch of data and without running any experiment. The formulation of DDSF with the sampled safe set, defined by (\ref{eq:CS_online}-\ref{eq:CS_offline}), is as follows:
\begin{subequations}\label{eq:DDSF}
\begin{align}\label{eq:DDSF_cost}
    \underset{\substack{\alpha(t),\bar{u}(t),\bar{y}(t)}}{\min}\>\>&\lVert \bar{u}_0(t)-u_{l}(t)\rVert_R^2 \\\label{eq:Hankel}
    \text{s.t.}\quad&\begin{bmatrix}\bar{u}(t)\\
    \bar{y}(t)\end{bmatrix}=\begin{bmatrix}H_{L}(u^d)\\H_{L}(y^d)\end{bmatrix}\alpha(t),\\\label{eq:initial_Condition}
    &\begin{bmatrix}\bar{u}_{[-T_{ini},-1]}(t)\\\bar{y}_{[-T_{ini},-1]}(t)\end{bmatrix}=\begin{bmatrix}u_{[t-T_{ini},t-1]}\\y_{[t-T_{ini},t-1]}\end{bmatrix},\\\label{eq:Final}
    &\bar{\xi}^{N-1}\in\Xi_f,\\\label{eq:I-O C}
    &\bar{u}_k(t)\in\mathcal{U},\>\>\bar{y}_k(t)\in\mathcal{Y},\>\>k\in\{0,\hdots,N-1\}.
\end{align}
\end{subequations}
\begin{figure}[h]
    \centering
    \includegraphics[width=0.9\linewidth]{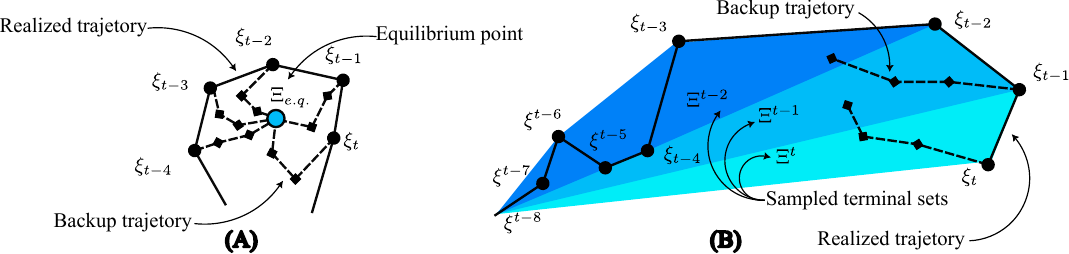}
    \caption{A visualization of the final set is shown. The solid black line shows the realized system trajectory. The dashed lines are the system's backup trajectories at time $t$. The blue polytopes show the convex hulls provided by extended states' realized trajectory. (A) The final set is the system's equilibrium point as proposed in \citep{bajelani2023data}. (B) The terminal set is expanded by the algorithm (\ref{algorithm: DDSF_online}) at times $t-2$, $t-1$, and $t$.}
    \label{fig: Extended_Sample_Safe_Set}
\end{figure}

By minimizing the cost function (\ref{eq:DDSF_cost}), the first input $u_{0}(t)$ becomes the nearest safe input to the potentially unsafe input $u_{l}$ while adhering to constraints (\ref{eq:Hankel}-\ref{eq:I-O C}). The implicit model for prediction, driven by measured raw data (\ref{equ: Input-output_data}), adheres to ($\ref{eq:Hankel}$) based on Theorem (\ref{Fundamental Lemma}). The optimization problem's initial condition is fixed using ($\ref{eq:initial_Condition}$), considering past input-output measurements. $\Xi_f$ in ($\ref{eq:Final}$) is defined as the terminal set for recursive feasibility and building backup trajectories, representing the sampled safe set calculated by the algorithm (\ref{algorithm: DDSF_online}) at time $t$ as $\Xi_f^{t-1}$ or the algorithm (\ref{algorithm: DDSF_offline}) at iteration $K$ as $\Xi_f^{K}$. For $t=1$ or $l=1$, $\Xi^{0}_f$ is assumed to be a known equilibrium point for the system (\ref{equ: true system}). Additionally, we consider that input-output constraints are defined by ($\ref{eq:I-O C}$). The following assumptions proposed in \citep{bajelani2023data} are also adopted.

\begin{Assumption}[Prediction Horizon Length] \label{Assumption: I}
    The prediction horizon $N$ is greater than $ T_{\text{ini}} \geq l(A, C)$.
\end{Assumption}
\begin{Assumption}[Persistent Excitation] \label{Assumption: II}
    The stacked Hankel matrix $[H_L(u^d) H_L(y^d)]$ is PE of order $L = N+2T_{\text{ini}}$ in the sense of definition (\ref{Def: PE condition}).
\end{Assumption}
\begin{Assumption}[Terminal Safe Set] \label{Assumption: III} The equilibrium point of the system (\ref{equ: true system}) defined in Definition (\ref{def: Equilibrium point.}) is known and used as the initial safe set of algorithms (\ref{algorithm: DDSF_online}-\ref{algorithm: DDSF_offline}).
\end{Assumption}
\begin{theorem}[Recursive Feasibility of DDSF with sampled safe set - online algorithm] \label{theorem: Recursive Feasibility}
Let assumptions (\ref{Assumption: I}-\ref{Assumption: III}) hold, $T_{\text{ini}} \geq \underline{l}$, $\Xi^{t_0}_f$ be a non-empty set. Then, the DDSF optimization problem (\ref{eq:DDSF}) is feasible for all $t>t_0$, if it is feasible at $t = t_0$.
\end{theorem}
{\bf Proof}. At $t=t_0$ the solution of optimization problem (\ref{eq:DDSF}) gives an input-output trajectory from the initial condition $\xi_{t=t_0}$ to the terminal set $\Xi_f^{t=t_0-1}$, as it is assumed the problem (\ref{eq:DDSF}) is feasible at $t=t_0$. At $t=t_0+1$, a new backup trajectory is calculated and the terminal set is expanded to  $\Xi_f^{t=t_0}$ by measuring $\xi_{t=t_0}$. Based on (\ref{subsets}), the terminal safe set is expanded implying $\Xi^{t=t_0-1} \subseteq \Xi^{t=t_0}$. Because the feasible set of the system (\ref{equ: true system}) is a convex set, if follows that the convex hull of all the realized extended states (or backup trajectories) is a subset of this set. If a system's trajectory enters this set, it will remain there permanently under the policy of (\ref{eq:DDSF}), thus demonstrating that all sampled safe sets are invariant and a subset of the feasible set. Therefore, by relying on induction, we can conclude that the problem (\ref{eq:DDSF}) remains feasible for all $t>t_0$.

\begin{Remark}
An under-approximation of the safe set may be necessary to expedite the optimization problem (\ref{eq:DDSF}) for higher-order systems. Specifically, when the system's lag is overestimated, the solution yields polytopes of extremely high dimensions, as the number of constraints depends on the estimation of the system's lag.
\end{Remark}

\section{Simulation Example} \label{Simulation}

To demonstrate the effectiveness of the proposed method, a time-delay second-order system is considered, as presented in (\ref{example: equ}). This system allows us to visualize the sampled terminal sets.
\begin{equation} \label{example: equ}
        x(t+1) = \left[\begin{array}{cc}
        1& -0.1 \\
        0& 1
        \end{array}\right] x(t) + \left[\begin{array}{c}
        0 \\
        0.1
        \end{array}\right]  u(t-\tau_d), \quad y(t) = \left[\begin{array}{cc}
        1& 0        \end{array}\right] x(t).       
\end{equation}
where the input-output constraints are assumed as follows,
\begin{subequations} \label{equ: I-O constranits-example} 
    \begin{align}   
        u(t) \in \mathcal{U} &:= \{ u \in \mathbb{R} \mid -1 \leq u \leq +1 \}, \\
        \quad y(t) \in \mathcal{Y} &:= \{ y \in \mathbb{R} \mid -1 \leq y \leq +1\}.
    \end{align}   
\end{subequations} 
To show the effect of the sampled safe set by the online and offline set expansion algorithms (\ref{algorithm: DDSF_online}-\ref{algorithm: DDSF_offline}) and the overestimation of the system's lag, two case studies are investigated. The impact of the sampled terminal set on the performance of a DDSF is demonstrated in section (\ref{First study}) for three safe sets, equilibrium point, and safe sets provided by online and offline algorithms when the system is subjected to delayed input. In section (\ref{Second study}), it is assumed that there is no delayed input but the effect of overestimating the system's lag is illustrated. Note that DDSFs are designed to keep input-output trajectories safe regardless of the control input; hence, we utilized unsafe control inputs to destabilize the system in this section. The oscillatory behavior of the output in the safe set, shown in light green, results from the periodic learning inputs and the correction provided by the DDSF policy.
\begin{figure} [h] 
    \centering
    \includegraphics[width=0.8\linewidth]{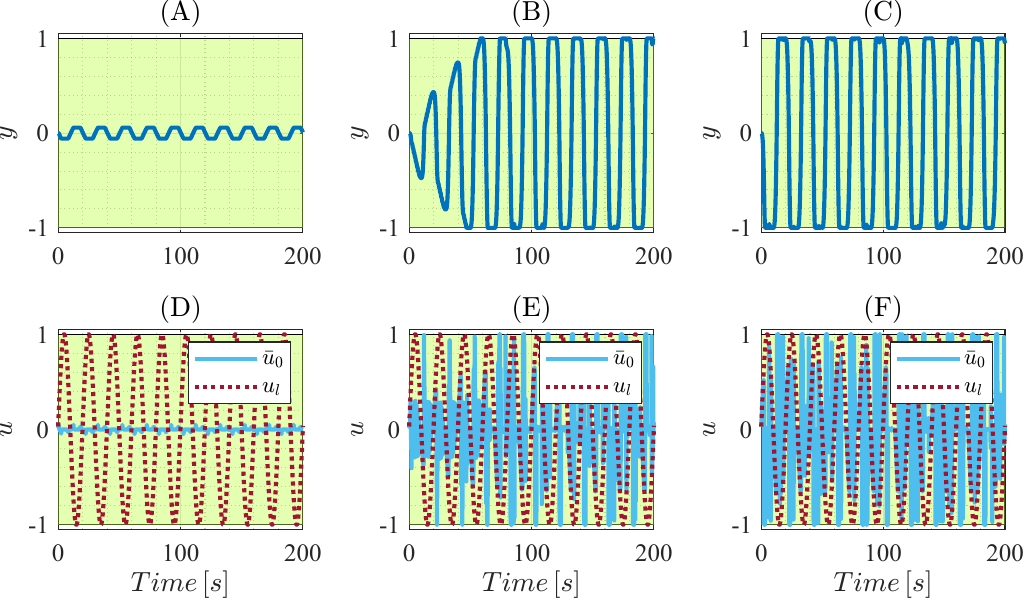}
    \caption{First case study: Input-output trajectories of system (\ref{example: equ}) under the policy of ($\ref{eq:DDSF}$) for three different safe sets: the system's equilibrium point, subplots (A) and (D), the sampled safe set computed by the online algorithm (\ref{algorithm: DDSF_online}), subplots (B) and (E), and the sampled safe set computed by the offline algorithm (\ref{algorithm: DDSF_offline}), subplots (C) and (F).}
    \label{fig: DDSF_Inputs_outputs}
\end{figure}
\subsection{First study: Sampled safe sets vs Equilibrium point safe set} \label{First study}
\begin{figure} [h] 
    \centering
    \includegraphics[width=0.85\linewidth]{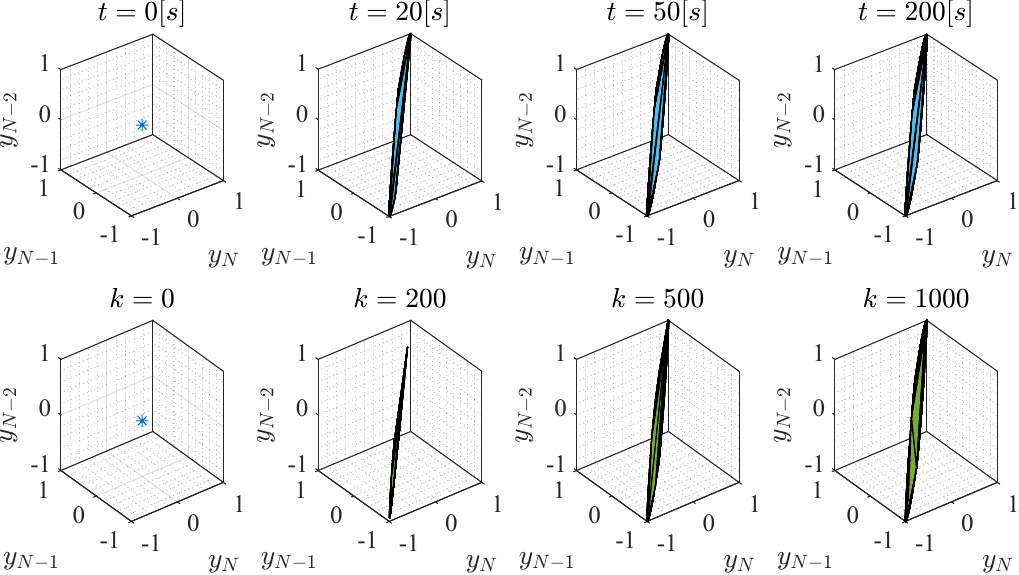}
    \caption{First case study: The sampled terminal safe set at four different times $t=0 [s]$, $t=20 [s]$, $t=50 [s]$, $t=200 [s]$ computed by the online algorithm (\ref{algorithm: DDSF_online}), and the sampled terminal safe set (\ref{algorithm: DDSF_online}) at four different iterations $k=0$, $k=200$, $k=500$, $k=1000$ computed by the offline algorithm (\ref{algorithm: DDSF_offline}).}
    \label{fig: Evolution_set_online_and_offline}
\end{figure}

For this study, a short prediction horizon $N_{p}=6$, dead-time $\tau_d=1$, and the corresponding known system's lag $T_{ini}=3$ are selected. It is also assumed that the learning input is a sinusoidal signal bounded in $[-1,+1]$. The equilibrium point $(u^s,y^s)=(0,0)$ is used as the initial terminal safe set, and algorithms (\ref{algorithm: DDSF_online}-\ref{algorithm: DDSF_offline}) are implemented using the CasADi toolbox \citep{Andersson2019}. The learning and safe inputs, along with the system's response for terminal sets are illustrated in Figure (\ref{fig: DDSF_Inputs_outputs}). It is evident that employing the sampled safe set as the terminal condition of (\ref{eq:DDSF}) reduces the conservatism of DDSF. Furthermore, it can be seen that at time $t=100 \, [s]$, the online approach is capable of reaching the boundary of the defined safe set. However, the convergence of the algorithms must be determined by the convergence of the safe set. The final safe sets for the online (\ref{algorithm: DDSF_online}) and offline (\ref{algorithm: DDSF_offline}) algorithms are illustrated in Figure (\ref{fig: Evolution_set_online_and_offline}) at different times and iterations. It must be noted that the final learned safe set provided by algorithms (\ref{algorithm: DDSF_online}) and (\ref{algorithm: DDSF_offline}) are the same. In other words, both the online (\ref{algorithm: DDSF_online}) and offline (\ref{algorithm: DDSF_offline}) algorithms yield the same safety filters after expanding the safe set.
\subsection{Second study: Overestimation of system's lag} \label{Second study}
\begin{figure}[h] 
    \centering
    \includegraphics[width=0.8\linewidth]{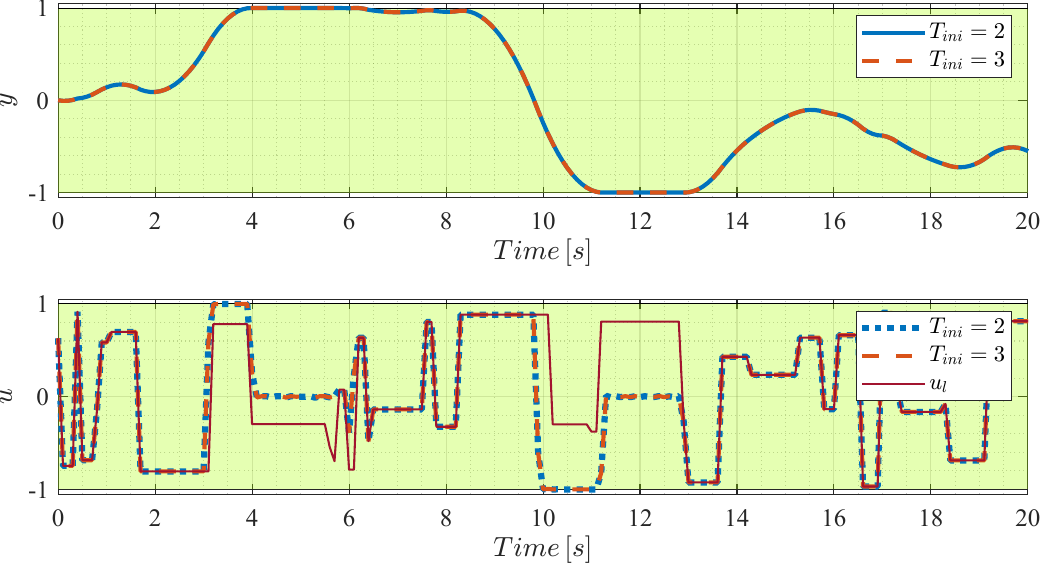}
    \caption{Second case study: Input-output trajectories of system (\ref{example: equ}) under the policy of ($\ref{eq:DDSF}$) for the exact and overestimated system's lags, $T_{\text{ini}}=2$ and $T_{\text{ini}}=3$, respectively.}
    \label{fig: Input_output_different_T_ini}
\end{figure}

In this part, it is assumed that $\tau_{d}=0$, the algorithm (\ref{algorithm: DDSF_offline}) is employed for $T_{ini} \in \{2,3\}$ and the learning input is a PRBS signal. The input-output behavior is depicted in Figure (\ref{fig: Input_output_different_T_ini}). It is clear that by overestimating the system's lag, the safety and performance of DDSF have not been affected as the input-output behavior is the same.
\vspace{-1cm}
\section{Conclusion}
To alleviate the conservatism associated with the Data-Driven Safety Filter (DDSF), this paper introduced online and offline set expansion algorithms based on sampled data. Specifically, we demonstrated the use of input-output data to compute safe sets in the form of extended state and backup trajectories. With these contributions, all stages of the design of a data-driven safety filter, from the raw data to the safety filter and safe sets, are purely data-driven, with no explicit modeling requirement. Additionally, by introducing an augmented system and treating delayed inputs as internal states, we illustrated the applicability of DDSF to unknown time-delay systems, leveraging sufficiently long past input-output data. Furthermore, we showed that safe sets can be computed offline using an implicit model derived from limited measured data. Note that if the system's lag is known, computing these sets becomes a straightforward task using backward reachable sets from an exact model. However, if this model is not exact, characterizing the impact of parametric or unmodeled uncertainty on the size and shape of the safe set is challenging. Due to the multi-step prediction property of the data-driven method proposed here, the effect of measurement noise on the computed safe set is direct. A potential avenue for future research involves exploring the impact of noise-polluted measurements and disturbances on the estimation of safe sets and the prediction of backup trajectories. Furthermore, the application of this methodology to controllers could be investigated, wherein backup trajectories and expanded terminal sets could be employed to design the terminal component of a data-driven predictive controller.
\acks{We acknowledge the support of the Natural Sciences and Engineering
Research Council of Canada (NSERC) [RGPIN-2023-03660].}
\bibliography{sample}

\begin{thebibliography}{26}
\providecommand{\natexlab}[1]{#1}
\providecommand{\url}[1]{\texttt{#1}}
\expandafter\ifx\csname urlstyle\endcsname\relax
  \providecommand{\doi}[1]{doi: #1}\else
  \providecommand{\doi}{doi: \begingroup \urlstyle{rm}\Url}\fi

\bibitem[Ames et~al.(2019)Ames, Coogan, Egerstedt, Notomista, Sreenath, and Tabuada]{ames2019control}
Aaron~D Ames, Samuel Coogan, Magnus Egerstedt, Gennaro Notomista, Koushil Sreenath, and Paulo Tabuada.
\newblock Control barrier functions: Theory and applications.
\newblock In \emph{2019 18th European control conference (ECC)}, pages 3420--3431. IEEE, 2019.

\bibitem[Andersson et~al.(2019)Andersson, Gillis, Horn, Rawlings, and Diehl]{Andersson2019}
Joel A~E Andersson, Joris Gillis, Greg Horn, James~B Rawlings, and Moritz Diehl.
\newblock {CasADi} -- {A} software framework for nonlinear optimization and optimal control.
\newblock \emph{Mathematical Programming Computation}, 11\penalty0 (1):\penalty0 1--36, 2019.
\newblock \doi{10.1007/s12532-018-0139-4}.

\bibitem[Bajelani and van Heusden(2023)]{bajelani2023data}
Mohammad Bajelani and Klaske van Heusden.
\newblock Data-driven safety filter: An input-output perspective.
\newblock \emph{arXiv preprint arXiv:2309.00189}, 2023.

\bibitem[Bansal et~al.(2017)Bansal, Chen, Herbert, and Tomlin]{bansal2017hamilton}
Somil Bansal, Mo~Chen, Sylvia Herbert, and Claire~J Tomlin.
\newblock {H}amilton-{J}acobi reachability: A brief overview and recent advances.
\newblock In \emph{2017 IEEE 56th Annual Conference on Decision and Control (CDC)}, pages 2242--2253. IEEE, 2017.

\bibitem[Berberich et~al.(2020{\natexlab{a}})Berberich, K{\"o}hler, M{\"u}ller, and Allg{\"o}wer]{berberich2020data}
Julian Berberich, Johannes K{\"o}hler, Matthias~A M{\"u}ller, and Frank Allg{\"o}wer.
\newblock Data-driven model predictive control with stability and robustness guarantees.
\newblock \emph{IEEE Transactions on Automatic Control}, 66\penalty0 (4):\penalty0 1702--1717, 2020{\natexlab{a}}.

\bibitem[Berberich et~al.(2020{\natexlab{b}})Berberich, K{\"o}hler, M{\"u}ller, and Allg{\"o}wer]{berberich2020robust}
Julian Berberich, Johannes K{\"o}hler, Matthias~A M{\"u}ller, and Frank Allg{\"o}wer.
\newblock Robust constraint satisfaction in data-driven {MPC}.
\newblock In \emph{2020 59th IEEE Conference on Decision and Control (CDC)}, pages 1260--1267. IEEE, 2020{\natexlab{b}}.

\bibitem[Berberich et~al.(2020{\natexlab{c}})Berberich, Köhler, Müller, and Allgöwer]{BERBERICH20206923}
Julian Berberich, Johannes Köhler, Matthias~A. Müller, and Frank Allgöwer.
\newblock Data-driven tracking mpc for changing setpoints.
\newblock \emph{IFAC-PapersOnLine}, 53\penalty0 (2):\penalty0 6923--6930, 2020{\natexlab{c}}.
\newblock ISSN 2405-8963.
\newblock 21st IFAC World Congress.

\bibitem[Berberich et~al.(2021)Berberich, K{\"o}hler, M{\"u}ller, and Allg{\"o}wer]{berberich2021design}
Julian Berberich, Johannes K{\"o}hler, Matthias~A M{\"u}ller, and Frank Allg{\"o}wer.
\newblock On the design of terminal ingredients for data-driven {MPC}.
\newblock \emph{IFAC-PapersOnLine}, 54\penalty0 (6):\penalty0 257--263, 2021.

\bibitem[Brunke et~al.(2022)Brunke, Greeff, Hall, Yuan, Zhou, Panerati, and Schoellig]{brunke2022safe}
Lukas Brunke, Melissa Greeff, Adam~W Hall, Zhaocong Yuan, Siqi Zhou, Jacopo Panerati, and Angela~P Schoellig.
\newblock Safe learning in robotics: From learning-based control to safe reinforcement learning.
\newblock \emph{Annual Review of Control, Robotics, and Autonomous Systems}, 5:\penalty0 411--444, 2022.

\bibitem[Coulson et~al.(2019)Coulson, Lygeros, and D{\"o}rfler]{coulson2019data}
Jeremy Coulson, John Lygeros, and Florian D{\"o}rfler.
\newblock Data-enabled predictive control: In the shallows of the {DeePC}.
\newblock In \emph{2019 18th European Control Conference (ECC)}, pages 307--312. IEEE, 2019.

\bibitem[Dörfler(2023)]{10317633}
Florian Dörfler.
\newblock Data-driven control: Part two of two: Hot take: Why not go with models?
\newblock \emph{IEEE Control Systems Magazine}, 43\penalty0 (6):\penalty0 27--31, 2023.

\bibitem[Herbert et~al.(2021)Herbert, Choi, Sanjeev, Gibson, Sreenath, and Tomlin]{herbert2021scalable}
Sylvia Herbert, Jason~J Choi, Suvansh Sanjeev, Marsalis Gibson, Koushil Sreenath, and Claire~J Tomlin.
\newblock Scalable learning of safety guarantees for autonomous systems using {H}amilton-{J}acobi reachability.
\newblock In \emph{2021 IEEE International Conference on Robotics and Automation (ICRA)}, pages 5914--5920. IEEE, 2021.

\bibitem[Hewing et~al.(2020)Hewing, Wabersich, Menner, and Zeilinger]{hewing2020learning}
Lukas Hewing, Kim~P Wabersich, Marcel Menner, and Melanie~N Zeilinger.
\newblock Learning-based model predictive control: Toward safe learning in control.
\newblock \emph{Annual Review of Control, Robotics, and Autonomous Systems}, 3:\penalty0 269--296, 2020.

\bibitem[Hsu et~al.(2023)Hsu, Choi, Amin, Tomlin, McWherter, and Piedmonte]{hsu2023towards}
Ting-Wei Hsu, Jason~J Choi, Divyang Amin, Claire Tomlin, Shaun~C McWherter, and Michael Piedmonte.
\newblock Towards flight envelope protection for the nasa tiltwing evtol flight mode transition using {H}amilton--{J}acobi reachability.
\newblock \emph{Journal of the American Helicopter Society}, 2023.

\bibitem[K{\"o}hler et~al.(2022)K{\"o}hler, Wabersich, Berberich, and Zeilinger]{kohler2022state}
Johannes K{\"o}hler, Kim~P Wabersich, Julian Berberich, and Melanie~N Zeilinger.
\newblock State space models vs. multi-step predictors in predictive control: Are state space models complicating safe data-driven designs?
\newblock In \emph{2022 IEEE 61st Conference on Decision and Control (CDC)}, pages 491--498. IEEE, 2022.

\bibitem[Markovsky and D{\"o}rfler(2021)]{markovsky2021behavioral}
Ivan Markovsky and Florian D{\"o}rfler.
\newblock Behavioral systems theory in data-driven analysis, signal processing, and control.
\newblock \emph{Annual Reviews in Control}, 52:\penalty0 42--64, 2021.

\bibitem[Markovsky et~al.(2006)Markovsky, Willems, Van~Huffel, and De~Moor]{markovsky2006exact}
Ivan Markovsky, Jan~C Willems, Sabine Van~Huffel, and Bart De~Moor.
\newblock \emph{Exact and approximate modeling of linear systems: A behavioral approach}.
\newblock SIAM, 2006.

\bibitem[Molnar and Ames(2023)]{molnar2023safety}
Tamas~G Molnar and Aaron~D Ames.
\newblock Safety-critical control with bounded inputs via reduced order models.
\newblock \emph{arXiv preprint arXiv:2303.03247}, 2023.

\bibitem[Rosolia and Borrelli(2017{\natexlab{a}})]{rosolia2017learning_1}
Ugo Rosolia and Francesco Borrelli.
\newblock Learning model predictive control for iterative tasks: A computationally efficient approach for linear system.
\newblock \emph{IFAC-PapersOnLine}, 50\penalty0 (1):\penalty0 3142--3147, 2017{\natexlab{a}}.

\bibitem[Rosolia and Borrelli(2017{\natexlab{b}})]{rosolia2017learning_2}
Ugo Rosolia and Francesco Borrelli.
\newblock Learning model predictive control for iterative tasks. a data-driven control framework.
\newblock \emph{IEEE Transactions on Automatic Control}, 63\penalty0 (7):\penalty0 1883--1896, 2017{\natexlab{b}}.

\bibitem[Tearle et~al.(2021)Tearle, Wabersich, Carron, and Zeilinger]{tearle2021predictive}
Ben Tearle, Kim~P Wabersich, Andrea Carron, and Melanie~N Zeilinger.
\newblock A predictive safety filter for learning-based racing control.
\newblock \emph{IEEE Robotics and Automation Letters}, 6\penalty0 (4):\penalty0 7635--7642, 2021.

\bibitem[Wabersich and Zeilinger(2018)]{wabersich2018linear}
Kim~P Wabersich and Melanie~N Zeilinger.
\newblock Linear model predictive safety certification for learning-based control.
\newblock In \emph{2018 IEEE Conference on Decision and Control (CDC)}, pages 7130--7135. IEEE, 2018.

\bibitem[Wabersich et~al.(2023)Wabersich, Taylor, Choi, Sreenath, Tomlin, Ames, and Zeilinger]{wabersich2023data}
Kim~P Wabersich, Andrew~J Taylor, Jason~J Choi, Koushil Sreenath, Claire~J Tomlin, Aaron~D Ames, and Melanie~N Zeilinger.
\newblock Data-driven safety filters: {H}amilton-{J}acobi reachability, control barrier functions, and predictive methods for uncertain systems.
\newblock \emph{IEEE Control Systems Magazine}, 43\penalty0 (5):\penalty0 137--177, 2023.

\bibitem[Wabersich and Zeilinger(2021)]{wabersich2021predictive}
Kim~Peter Wabersich and Melanie~N Zeilinger.
\newblock A predictive safety filter for learning-based control of constrained nonlinear dynamical systems.
\newblock \emph{Automatica}, 129:\penalty0 109597, 2021.

\bibitem[Wang et~al.(2018)Wang, Theodorou, and Egerstedt]{8460471}
Li~Wang, Evangelos~A. Theodorou, and Magnus Egerstedt.
\newblock Safe learning of quadrotor dynamics using barrier certificates.
\newblock In \emph{2018 IEEE International Conference on Robotics and Automation (ICRA)}, pages 2460--2465, 2018.

\bibitem[Willems et~al.(2005)Willems, Rapisarda, Markovsky, and De~Moor]{willems2005note}
Jan~C Willems, Paolo Rapisarda, Ivan Markovsky, and Bart~LM De~Moor.
\newblock A note on persistency of excitation.
\newblock \emph{Systems \& Control Letters}, 54\penalty0 (4):\penalty0 325--329, 2005.

\end{thebibliography}
\end{document}